\documentclass[openany,oneside,12pt,a4paper,onecolumn]{article}

\def\seCtion#1{\section{#1} \setcounter{equation}{0}}
\renewcommand\theequation{\ifnum\value{section}>0{\thesection.
\arabic{equation}}\fi}
\newcommand{\be}{\begin{equation}}
\newcommand{\ee}{\end{equation}}
\newcommand{\bea}{\begin{eqnarray}}
\newcommand{\eea}{\end{eqnarray}}

\usepackage{graphics}
\usepackage{latexsym}
\usepackage{array}
\usepackage[activeacute,english]{babel}
\usepackage[latin1]{inputenc}
\usepackage[dvips]{graphicx}
\usepackage{color}

\begin{document}

\pagestyle{empty}
\begin{flushright}
FIUN-GCP-11/01
\end{flushright}

\begin{center}
{\Large \bf Cosmological constant in a quantum fluid model}\\

\vspace{0.5cm}

{\bf J. A. S\'anchez-Monroy \footnote{jasanchezm@unal.edu.co.} and
C. J. Quimbay \footnote{cjquimbayh@unal.edu.co.,
associate researcher of CIF, Bogot\'a, Colombia.}}\\
{\it{Departamento de F\'{\i}sica}, Universidad Nacional de
Colombia\\
Ciudad Universitaria, Bogot\'a, Colombia}\\
\vspace{0.5cm} September 10, 2011
\end{center}

\vspace{0.8cm}

\begin{abstract}

Possible analogies between vacuum state and quantum fluid provide
a model to study vacuum energy density induced by thermal
corrections, space-time curvature, boundary conditions and quantum
back-reaction. We find that vacuum energy density in this quantum
fluid model is not naturally of the order of the matter energy
density. We show how higher-order corrections in quantum
back-reaction can also contribute to vacuum energy density, and
how the cosmological expansion is a manifestation of an universe
out of mechanical equilibrium. This last fact implies that simple
thermodynamic arguments are not enough to explain the cosmological
constant problem due to the calculation of the associated vacuum
energy density requires first the knowing of the underlying
microscopic physics of vacuum.
\end{abstract}

\pagestyle{plain}


\seCtion{Introduction}

The idea of building analogue models between laboratory physics
and cosmological phenomena has attracted a great interest
\cite{Ralf3}. It has been also considered analogies between
particle physics and condensed matter systems \cite{Voll,Xiao}.
Possible analogies between many-body quantum mechanics and
Relativistic Quantum Field Theory (RQFT) are based on considering
the gauge bosons and Dirac fermions as quasiparticle excitations
of a quantum liquid. Moreover for low-energy phenomena it is
possible to consider the Standard Model of Particle Physics (SM)
and the General Relativity Theory as effective theories which
emerge from fermion zero modes of the quantum liquid vacuum state
\cite{ca4,ca5}. In particular the cosmological constant problem
has been studied under this scheme \cite{ca4}-\cite{epove}.
\par
Physics for a weakly interacting Bose gas can be modeled as the
ground state of an interacting boson system plus a set of
excitations (quasiparticles). A phenomenological description at
low energy of a Bose quantum liquid can be performed by means of
Bogoliubov transformations. The ground state $|0\rangle$ is such
that the annihilation operator of quasiparticles
$\hat{\alpha}_{\mathbf{p}}$ annihilates the ground state
\cite{Xiao,ca5} $\hat{\alpha}_{\mathbf{p}}|0\rangle=0$, in a
similar way as the annihilation operator of particles annihilates
the RQFT vacuum state. The effect of Bogoliubov transformations in
a Fermi liquid is similar to the one at a Bose quantum liquid but
in this case the ground state is a Bardeen-Cooper-Schrieffer (BCS)
state \cite{ca5,SPH3}. The RQFT vacuum state differs from the
ground state of a quantum mechanical system due to the fact that
the ground state of this last system can scatter particles while
the RQFT vacuum state can not \cite{rugh}. Quantum fluctuations of
vacuum can not scatter quasiparticles because the homogeneous
vacuum state of a quantum liquid can not scatter them \cite{ca5}.
\par
The structure of quantum liquids remains known over a microscopic
(trans-Planckian) scale in contrast to what happens in the SM
where the structure is unknown. However starting from topological
properties of the SM one might suspect that the vacuum state of
this model in the unbroken electroweak symmetry has the same
universality class \cite{ca5} as the one of $^3$He-A. Since gauge
symmetry of a Hamiltonian describing a relativistic chiral
particles system is a property that can emerge from the Fermi
point universality class \cite{ca4,ca5}, it has been conjecture
that all bosons and fermions of the SM can emerge in the vicinity
of Fermi points \cite{ca4,ca5}.
\par
On the other hand, there exists a deep connection between vacuum
energy density and cosmological constant. Einstein introduced the
cosmological constant in the field equation for gravity with the
motivation that it could carry out a finite, closed, static
universe in which the energy density of matter determines geometry
\cite{ca}-\cite{ca1}. From observations \cite{ia} of SNIa combined
with CMB anisotropies \cite{WMAP2007} it has been suggested that
the expansion of universe is increasing in an accelerated manner.
The acceleration is driven by an unknown form of dark energy
having a relative density \cite{WMAP2007} of
$\Omega_{\Lambda}=0.726 ± 0.015$. For the dark energy density, the
observations imply a possible value for $\omega$ in the range
\cite{WMAP2007} $-0.14<1+\omega <0.05$, where $\omega$ is the
parameter that relates pressure $P$ and dark energy density $\rho$
in the equation of state $P=\omega \rho$. Although the nature of
dark energy is a complete mystery, the observations are in
agreement with the idea that dark energy could arises from a pure
cosmological constant term in the Einstein field equation. A
positive cosmological constant $\Lambda$ of magnitude
$\Lambda(G\hbar/c^3)\leq 10^{-123}$ can be associated with the
dark energy density by means of the Einstein field equation
\cite{ca1}. We remind that the cosmological constant problem is
related to establish what the dark energy origin is.
\par
The aim of this paper is to analyze some possible contributions of
vacuum state to cosmological constant in a model which is based on
considering the vacuum state as a quantum fluid. Several
contributions to the cosmological constant have been studied under
this model \cite{ca6,ca9}. One of these contributions is the
presence of matter in the universe. But since this contribution
depends only on the state equation of matter, it has been
suggested that the coincidence problem\footnote{The coincidence
problem consists of understanding why the present cosmological
observations show that the order of magnitude of vacuum energy
density is the same as the order of magnitude of matter density.}
is partially solved \cite{ca6}. However a direct calculation as
presented in section \ref{VEFFT} shows that the observed
cosmological constant is four order of magnitude higher than the
one induced by matter contribution. We find in section \ref{QBR}
that higher-order corrections to vacuum energy density depend on
the microscopic structure of the theory. Finally we conjecture in
section \ref{VEINE} that the cosmological expansion implies that
the universe is not in a mechanical equilibrium state and this
fact might be a meaningful effect for the cosmological constant
problem.


\seCtion{Vacuum state as a quantum fluid}\label{VSAQF}

Vacuum energy density in RQFT can be estimated from positive and
negative contributions. A positive contribution comes from
zero-point energy of bosonic fields while a negative contribution
comes from occupied negative energy levels in the Dirac sea
\cite{ca9}. For the case in which energy spectrum of particles is
massless $E\sim cp$, the energy density of quantum vacuum
$\rho_{\Lambda}$ can be expressed in terms of number of bosonic
$\nu_b$ and fermionic $\nu_f$ species such as \cite{ca6}
\begin{equation}\label{fluid4}
\rho_{\Lambda}=\frac{1}{2V}\sum_{b,\mathbf{p}}cp-\frac{1}{V}
\sum_{f,\mathbf{p}}cp \sim
\frac{1}{c^3}\left(\frac{1}{2}\nu_b-\nu_f\right)
E^4_{Planck}=\sqrt{-g}\left(\frac{1}{2}\nu_b-\nu_f\right)
E^4_{Planck},
\end{equation}
because the largest contribution comes from the high momenta. The
cut-off is provided by the Planck energy scale $E_{Planck} \sim
10^{19}$ GeV. The vacuum energy density obtained is too large
respect to the observed value.
\par
On the other hand the vacuum of a quantum liquid receives
contributions from the trans-Planckian and sub-Planckian degrees
of freedom. These degrees of freedom describe the interacting and
correlated system of atoms in a real liquid. The calculation of an
exact energy associated with the many-body wave function which
describes the ground state of this real liquid has been performed
\cite{ca9}. Additionally the effect of ground state energies for
weakly interacting Bose and Fermi liquids is known
\cite{Andersen}.
\par
An appropriate model for the cosmological constant needs to
satisfy the equation of state $P_{\Lambda}=-\rho_{\Lambda}$, where
$P_{\Lambda}$ is the pressure and $\rho_{\Lambda}$ is the energy
density of quantum vacuum. Owing to the ground state of quantum
liquids has associated thermodynamical relations at temperature
$T$ which lead to the necessary equation of state, vacuum state
can be considered as a quantum fluid. This quantum fluid is
constituted by weakly interacting Bose and Fermi liquids. The
pressure $P$ for the quantum fluid at $T=0$ can be indistinctly
determined through three different thermodynamic potentials:
Helmholtz free energy, internal energy and grand potential. Using
the grand potential $\Omega(T,V,\mu)$, the pressure can be written
as \cite{Qfticmp}
\begin{eqnarray}\label{pre1}
P=-\frac{1}{V}\langle 0|\hat{H}-\mu \hat{N}|0\rangle \equiv
-\tilde{\epsilon},
\end{eqnarray}
where $|0\rangle$ is the ground state of the quantum fluid,
$\hat{H}$ is the Hamiltonian operator, $\hat{N}$ is the particle
number, $\mu$ is the chemical potential and $V$ is the volume. We
can identify $\tilde{\epsilon}$ as the dark energy density due to
the fact that the expression (\ref{pre1}) is clearly the equation
of state for dark energy. Thus the term $\hat{H}-\mu \hat{N}$
takes into account the ligature on particle number and the grand
potential is independent from the choice of energy reference
\cite{ca6}.
\par
On this model it has been considered that if the universe is in a
thermodynamic equilibrium state, in absence of an external
environment (it means a vanishing external pressure), the exact
nullification of vacuum energy $\tilde{\epsilon}$ occurs without
any special fine-tuning. This last fact can be obtained because
the thermodynamic relation carries on the whole equilibrium
\cite{ca4,ca5,ca6,ca9}. The last issue is true if we consider that
for a quantum liquid in mechanical equilibrium the internal
pressure is equal to the external one. But, as discussed below,
internal pressure is not only originated by vacuum pressure.

Since dark energy can be considered as a perfect fluid, the role
that the observer speed has over the dark energy density should be
analyzed. To do that, we remind that the stress-energy tensor
$T^{\mu \nu}$ for a perfect fluid is given by
\begin{equation}\label{fluid}
T^{\mu \nu}=-Pg^{\mu \nu}+(\rho +P)u^{\mu}u^{\nu},
\end{equation}
where $g_{\mu \nu}$ is the metric tensor, $u^{\nu}$ is the
four-velocity and $\rho$ is the dark energy density. For
simplicity, the universe is considered into a special relativity
framework \cite{Vol1}, i.e. in absence of gravity $G=0$. In a
general coordinate frame, energy density $T^M_{0 0}$ and momentum
density $T^M_{0 i}$ are \cite{Vol1}
\begin{equation}\label{fluid2}
T^M_{0 0}=\gamma ^2\left(\rho^M+\frac{v^2}{c^2}P^M\right), \ \ \ \
\ \ \ \ \ \ \ \ T^M_{0 i}=\gamma ^2\left(\rho^M+P^M\right)v_i,
\end{equation}
where $v$ is the speed respect to the rest frame of the fluid and
$\gamma=1/\sqrt{1-v^2/c^2}$. Since the cosmological constant
satisfies $P_{\Lambda}=-\rho_{\Lambda}$, the stress-energy tensor
satisfies $T^{\mu \nu}_\Lambda=\rho_{\Lambda} g^{\mu \nu}$ for any
general coordinate frame. For a dark energy model in which the
equation of state is $P\neq-\rho$, the energy density $T_{00}$
depends on the relative velocity between the rest frame of the
fluid and the observer frame. This means that if dark energy is
modeled by a fluid, satisfying an equation of state given by
$P\neq-\rho$, dark energy density depends on the observer speed.

The thermodynamic behavior of a relativistic perfect simple fluid
that obeys an equation of state of the form $P=\omega \rho$ has
been studied from a general perspective \cite{Lima1,Lima2}. For
the particular case of an adiabatic expansion of the universe and
modeling the dark energy with $\omega =-1$, the pressure is
negative. For this case, the thermodynamic work is being
continuously done on each volume element by the rest of the
universe, and it is possible to conclude that \cite{Lima1,Lima2}:
(i) dark energy density remains constant, (ii) entropy is null,
and (iii) temperature satisfies the relation $T\propto V$.
Additionally, since work is done on the system, dark energy and
temperature of the dark energy component grow during the evolution
of the universe \cite{Lima2}. If the universe is dominated by dark
energy then the expansion will be forever and the universe will
become increasingly hot \cite{Lima2}.


\seCtion{Contributions to the cosmological constant}

On this section we will analyze the possible contributions of
vacuum state to cosmological constant. A complete discussion about
the meaning of these possible contributions, considering different
scenarios for the quantum fluid, has been performed in Refs.
\cite{ca4}-\cite{ca9}. The vanishing value of vacuum energy
obtained from equilibrium condition of quantum liquids is
perturbed by different scenarios.
\par


\subsection{Vacuum energy from finite temperature}\label{VEFFT}

The external pressure $P_{E}$ in a thermodynamical system vanishes
if there is no external force. If this system is in equilibrium
then the internal pressure $P_{I}$ vanishes too. It is possible to
think in an universe where pressure $P_M$, which is originated by
quasiparticles (playing the role of matter), is compensated by a
negative vacuum pressure $P_{\Lambda}$, in such a way that the
internal pressure of the universe is $P_{I}=P_{\Lambda}+P_M=0$.
For an universe in equilibrium and assuming the absence of the
external environment, the external pressure satisfies
\cite{ca4}-\cite{ca9} $P_{E}=P_{I}=0$. Considering that the
universe is composed by several species of matter $i$ and the
equation of state for a specie $i$ is $P_i= \omega_i \rho_i$, we
obtain
\begin{equation}
P_{E}=P_{\Lambda}+\sum_iP_i=P_{\Lambda}+\sum_i\omega_i \rho_i=0,
\end{equation}
implying that $P_{\Lambda}=-\sum_i\omega_i \rho_i$. Using the
equation of state for the cosmological constant
$P_{\Lambda}=-\rho_{\Lambda}$, we have that the vacuum energy
density is given by
\begin{equation}
\rho_{\Lambda}=\sum_i\omega_i \rho_i.
\end{equation}
Species of matter contributing to vacuum energy density are
baryonic matter, dark matter and hot relativistic matter (photons)
\begin{equation} \rho_{\Lambda}=\omega_{bm} \rho_{bm}+\omega_{dm}
\rho_{dm}+\omega_{rm} \rho_{rm}.
\end{equation}
For baryonic matter is known that $\omega_{bm}=0$, implying that
baryonic matter does not contribute to $\rho_{\Lambda}$. Since
dark matter is cold, its equation of state is
$P_{dm}=\omega_{dm}\rho_{dm}$. Cosmological bounds \cite{Muller}
suggest that $\omega_{dm}$ has a possible value in the range
$-1.50\times 10^{-6}<\omega_{dm}<1.13\times 10^{-5}$. So dark
matter contribution to $\rho_{\Lambda}$ can be depressed. Thus,
the only contribution to $\rho_{\Lambda}$ is due to hot
relativistic matter. This contribution can be seen as thermal
quantum corrections to the ground state energy density (vacuum
energy) of a weakly interacting Bose gas \cite{ca6,Andersen}. For
$T\ll 2\mu$, where the small evaporation can be neglected
exponentially, the quantum liquid can be considered in equilibrium
\cite{ca9}. The ground state energy density
$\tilde{\epsilon}_{rm}$, that includes a thermal quantum
correction coinciding with the Stefan-Boltzmann law, is given by
\begin{equation}
\tilde{\epsilon}_{rm}=\tilde{\epsilon}_{rm}(T=0)+\rho_{rm}=
\tilde{\epsilon}_{rm}(T=0)+\sqrt{-g}\frac{\pi^2k_B^4T^4}{30\hbar^3},
\end{equation}
where $\tilde{\epsilon}_{rm}(T=0)$ is the vacuum energy density at
zero temperature, $k_B$ is the Boltzmann constant and
$\sqrt{-g}=c^{-3}$, being $c$ the sound velocity
\cite{ca6,Andersen}. On this calculation we have considered the
fact that a photon has two polarizations. Since hot relativistic
matter satisfies the equation of state given by $P_{rm}=
1/3\rho_{rm}$, hot relativistic matter contributes to
$\rho_{\Lambda}$ as
\begin{equation}
\rho_\Lambda=\frac{1}{3}\rho_{rm}=\frac{1}{3}\sigma
T^4=\frac{1}{3}\frac{\pi^2k_B^4T^4}{15c^3\hbar^3}.
\end{equation}
Thus the energy density parameter associated with hot relativistic
matter is \cite{WMAP2007} $\Omega_{rm}\approx4.8(4)\times10^{-5}$.
Therefore the induced value for the vacuum energy density
parameter is
\begin{equation}
\Omega_\Lambda\approx 1.6\times10^{-5},
\end{equation}
which is a value four order of magnitude smaller than the
observed. This value is not in agreement with the result given in
current literature \cite{ca4,ca5,ca6,ca9}, where the coincidence
problem is solved. This calculation shows that this model is not
in agreement with experimental data.


\subsection{Vacuum energy due to a topological defect}

It has been suggested that a nonzero vacuum energy can be induced
by the inhomogeneity of vacuum in quantum liquids. As an example
of this idea, it was studied a special kind of topological defect
called texture \cite{ca4,ca5,ca6}. It was found that there is an
equivalence among the energy gradient of twisted texture in
$^3$He-A and the Riemann curvature of an effective space with time
independent metric \cite{ca5}. For this reason the vacuum energy
density induced by the texture is proportional to the curvature
$k$. As the universe is flat \cite{WMAP2007}, i. e. $k\approx 0$,
then the induced vacuum energy density vanishes. However, the
analogy with quantum liquids suggests naturally that the universe
is flat \cite{ca5,ca9}.


\subsection{Quantum back-reaction}\label{QBR}

As it was previously mentioned, it has been conjectured that all
the bosons and fermions of the SM emerge in the vicinity of Fermi
points. Particularly, in the vicinity of a Fermi point,
quasiparticles (which by analogy correspond to Dirac fermions) are
massless chiral fermions moving in gravitational and
electromagnetic effective fields generated by collective movement
of vacuum at low frequencies \cite{ca5}. On this scheme, the SM is
equivalent to an effective theory of quasiparticles in a quantum
liquid at low frequencies.
\par
We can study the analogy between SM particles and quasiparticles
at low frequency by considering the approach of quantum
back-reaction of dilute Bose-Einstein condensates which has been
recently studied \cite{Ralf3,Ralf1,Ralf2}. This approach is based
on in considering the classical quantities plus small quantum
fluctuations $\hat{\psi}=\psi_c+\delta\hat{\psi}$. Implications of
this approach at different levels have been studied accurately by
Schtzhold \cite{Ralf}. Particularly the study of the
time-dependent Gross-Pitaevskii equation reveals that the
perturbation (linear quantum fluctuation)
$\delta\hat{\psi}(\mathbf{r},t)$ corresponds to quasiparticles
(excitations). This perturbation is acting over the unperturbed
wave function of the condensed state
$\psi(\mathbf{r},t)=\sqrt{n(\mathbf{r})}e^{-i\mu t/\hbar}$, where
$n(\mathbf{r})$ is the equilibrium density of particles and $\mu$
the chemical potential of the unperturbed system \cite{Becidg}. On
this scheme of vacuum state as a quantum fluid, linear quantum
fluctuations (one-particle excitations) are describing the matter.
\par
We have a better approximation if we consider the Gross-Pitaevskii
equation given by
\begin{equation}
-\frac{\hbar^2}{2m}\nabla^2\hat{\Psi}(\mathbf{r},t)+
V(\mathbf{r})\hat{\Psi}(\mathbf{r},t)+g|\hat{\Psi}(\mathbf{r},t)|^2
\hat{\Psi}(\mathbf{r},t)=i\hbar\frac{\partial
\hat{\Psi}(\mathbf{r},t)}{\partial t},
\end{equation}
where $V(\mathbf{r})$ denotes the external one-particle potential
and $g$ an effective coupling constant. Here we have substituted
the wave function $\psi(\mathbf{x},t)$ by its corresponding full
field operator $\hat{\Psi}$. This operator can be represented in
terms of the particle-number-conserving mean-field ansatz as
\cite{Ralf3,Ralf1,Ralf2}
\begin{equation}
\hat{\Psi}=(\psi_c+\hat{\chi}+\hat{\zeta})\hat{A}\hat{N}^{-1/2},
\end{equation}
where $\psi_c$ is the order parameter, $\hat{\chi}$ is the
single-particle excitations and $\hat{\zeta}$ is the higher-order
corrections originated in multi-particle excitations and
correlations \cite{Ralf1,Ralf2}. Here single-particle means that
Fourier components of $\hat{\chi}$ are linear superpositions of
annihilation and creation operators of quasiparticles and
$\hat{N}=\hat{A}^\dag \hat{A}$ counts the total number of
particles \cite{Ralf1,Ralf2}.
\par
The vacuum pressure induced by single-particle excitations only
represent contributions of matter. At this respect we note that
higher-order corrections also contribute to ground-state energy,
as can be seen in the equations of motion coupled by $\psi_c$,
$\hat{\chi}$ and $\hat{\zeta}$ \cite{Ralf1,Ralf2}. Since the
higher-order corrections depend on microscopic details of the
interactions among the fundamental constituents of the quantum
liquid, microscopic physics of the system should be known for
obtaining the full contributions to the vacuum pressure.

The surface tension is another scenario that might induce a
cosmological constant for a universe in mechanical equilibrium.
This tension is provided by boundaries of the system
\cite{ca4,ca5,ca9}, meaning that the universe is bounded by a
surface in a $3$-dimensional space. However this scenario has not
been considered here.\par


\seCtion{Vacuum energy in non-equilibrium}\label{VEINE}

An universe in mechanical equilibrium implies that there are no
external forces acting over itself, i.e. an universe in this state
is not able to experiment a spontaneous change of state when it is
subjected to certain boundary conditions. As mentioned before, the
mechanical equilibrium of the universe occurs when the external
pressure is equal to the internal one $P_{E}=P_I$. Some
implications of this mechanical equilibrium can be analyzed by
thinking in a mechanical system in absence of gravity. Let us
considering first a gas into a cubic recipient of volume $V$
having a piston over one of its walls. If the external pressure
exerted by the piston is equal in magnitude, but in opposite
sense, to the pressure exerted by the gas on the wall then this
system is in mechanical equilibrium. Secondly, let a drop of
volume $V$ in a vacuum space. If there are no external forces
acting over the drop then the external pressure vanishes. Here the
drop will remain with its volume $V$ if the internal pressures are
canceled, otherwise the drop will expand.
\par
As the universe has been taken in absence of an external
environment, the external pressure vanishes $P_{E}=0$. For the
case in which the universe is not in a mechanical equilibrium
state then the internal pressure does not vanish. For this case
the expansion of the underlying quantum fluid is similar as a drop
which is expanding in absence of external forces. As a consequence
there exists a cosmological expansion of the universe. We
conjecture that the effect of an universe out of mechanical
equilibrium can be seen by means of a meaningful factor which
could contribute to vacuum energy density. In that case the
underlying microscopic physics should be known so that the energy
associated to vacuum can be calculated. In other words, if simple
thermodynamic arguments are not sufficient to calculate the vacuum
energy density in an universe out of mechanical equilibrium, it is
necessary to know what the underlying dynamics and structure of
the quantum fluid are in order to describe the vacuum state.
\par
The effect of a non-equilibrium vacuum state over the vacuum
energy has been studied in the context of a weakly interacting
Fermi gas approach using the BCS theory \cite{ca5}. On this
approach it has been established that for an universe in
mechanical equilibrium the vacuum state is not gravitating and the
cosmological constant vanishes \cite{ca5}. Likewise for the case
of an universe out of equilibrium the vacuum energy contributes to
the cosmological constant \cite{ca5}. Thus the dark energy density
does not match the vacuum energy density. For this reason, it is
necessary to re-escale the vacuum energy density by a factor that
depends on the underlying microscopic physics of the Fermi gas
\cite{ca5}.
\par
Now we will determine what the effects of a non-equilibrium vacuum
state are over the thermodynamic behavior of a relativistic
perfect simple fluid that obeys an equation of state of the form
$P=- \rho$ in a flat space-time. The energy-momentum conservation
law remains valid both for equilibrium and nonequilibrium, $T^{\mu
\nu};_{\nu}=0$, where a semicolon denotes a covariant derivative.
This leads to $\dot{\rho}=0$, where an overdot means a comoving
time derivative. Cosmological expansion is an adiabatic process,
nevertheless, when a process takes place between non-equilibrium
states, the definition of the entropy is an open problem which has
not been yet definitively solved \cite{Lebon}. However, we assume
the local equilibrium hypothesis and use the Gibbs-Duhem relation
\begin{equation}
G=U-TS+PV=\mu N,
\end{equation}
that together with the first law of thermodynamics, leads us to
write
\begin{equation}
nTd\sigma=d\rho-\frac{p+\rho}{n}dn,
\end{equation}
where $n$ is the particle number density, $\sigma$ is the specific
entropy (per particle) and $T$ is the temperature. After use the
relation $P=- \rho$ into the Gibbs law, we obtain
\begin{equation}
T\left(\frac{\partial p}{\partial
T}\right)_n=-n\left(\frac{\partial \rho}{\partial n}\right)_T.
\end{equation}
Due to we can assume that $n$ and $T$ are independent
thermodynamic variables, the last equation can driven to
\begin{equation}
\frac{\dot{T}}{T}=\left(\frac{\partial p}{\partial \rho}\right)_n
\frac{\dot{n}}{n}=-\frac{\dot{n}}{n}.
\end{equation}
After a straightforward integration we can obtain $Tn=$const.
Since $n$ scales with $V^{-1}$, where $V$ is the volumen, then
$TV=$const. Thus, for an universe in a non-equilibrium state, if
we assume the validity of the local equilibrium hypothesis and we
model the dark energy as a relativistic perfect simple fluid
satisfying $P=-\rho$, we find that the thermodynamic behavior of
this universe is the same to the one found in Refs.
\cite{Lima1}-\cite{Lima2} for the case of an universe in an
equilibrium state. We have mentioned this thermodynamic behavior
in the last part of section \ref{VSAQF}.


\seCtion{Conclusions}

A thermodynamic analysis of quantum vacuum which bears under a
model that considering the vacuum state as a quantum fluid carries
us to analyze different contributions to cosmological constant. We
have found that the vacuum energy density calculated under this
scheme is not of the order of energy density of matter. We have
found that the textures do not contribute to vacuum energy density
for a flat universe, while higher-order corrections in quantum
back-reaction do. Furthermore we have conjectured that the
cosmological expansion is a manifestation of an universe out of
mechanical equilibrium. This fact makes necessary to know the
underlying microscopic physics to be able to calculate the
associated vacuum energy and therefore there is no cancelation of
the cosmological constant. The last fact shows that simple
thermodynamic arguments are not enough to calculate vacuum energy
in contrast to what has been suggested in the literature.

\section*{Acknowledgments}

This work was supported by DIB under a research grant "Grupo de
Campos y Partículas". We are indebted to Rafael Hurtado and
Christian Rojas for stimulating discussions.

\end{document}